# Counterfactuals and Uncertainty-Based Explainable Paradigm for the Automated Detection and Segmentation of Renal Cysts in Computed Tomography Images: A Multi-Center Study

Zohaib Salahuddin[1], Abdalla Ibrahim[2,*], Sheng Kuang[1,*], Yousif Widaatalla[1], Razvan L. Miclea[3], Oliver Morin[4], Spencer Behr[4], Marnix P.M. Kop[5], Tom Marcelissen[6], Patricia Zondervan[5], Auke Jager[5], Philippe Lambin[1,3], Henry C Woodruff[1,3]

[1] The D-Lab, Department of Precision Medicine, GROW – School for Oncology and Reproduction, Maastricht University, Maastricht, The Netherlands
[2] Department of Radiology, Memorial Sloan Kettering Cancer Center, New York, USA
[3] Department of Radiology and Nuclear Medicine, GROW – School for Oncology and Reproduction, Maastricht University Medical Center+, Maastricht, The Netherlands
[4] Department of Radiation Oncology, University of California-San Francisco, USA
[5] Department of Radiology, Amsterdam UMC, University of Amsterdam, Amsterdam, The Netherlands
[6] Department of Urology, GROW Research Institute for Oncology and Reproduction, Maastricht University Medical Center, Maastricht, The Netherlands

## Abstract

Routine computed tomography (CT) scans often detect a wide range of renal cysts, some of which may be malignant. Early and precise localization of these cysts can significantly enhance patient outcomes and aid quantitative image analysis. Current segmentation methods, however, do not offer sufficient interpretability at the feature and pixel levels, emphasizing the necessity for an explainable framework that can detect and rectify model inaccuracies. We developed an interpretable segmentation framework, validated on a dataset of 568 patients and 1,468 cysts, from three different centers and an open-source dataset. This framework first isolated the kidney region and then performed 3D segmentation of the renal cysts. It also incorporated segmentation quality control through uncertainty estimation. A Variational Autoencoder Generative Adversarial Network (VAE-GAN) was employed to learn the latent representation of 3D input patches and reconstruct input images. Modifications in the latent representation using the gradient of the segmentation model generated counterfactual explanations for varying dice similarity coefficients (DSC). Radiomics features extracted from these counterfactual images, using a ground truth cyst mask, were analyzed to determine their correlation with segmentation performance. The framework achieved a DSC of 0.82 on two external test sets, with sensitivities of 0.95 and 0.97, respectively. The DSCs for the original and VAE-GAN reconstructed images for counterfactual image generation showed no significant differences. Counterfactual explanations highlighted how variations in cyst image features influence segmentation outcomes and showed model discrepancies. Radiomics features correlating positively and negatively with dice scores were identified. Using uncertainty estimates to flag segmentations with dice scores below 0.75 reduced the percentage of poor segmentations from 17% to 5%, when only 20% of the data was flagged for manual correction. The combination of counterfactual explanations and uncertainty maps provided a deeper understanding of the image features within the segmented renal cysts that lead to high uncertainty. The proposed segmentation framework not only achieved high segmentation accuracy but also increased interpretability regarding how image features impact segmentation performance. The proposed interpretability method can also be used for other segmentation problems.

---

* These authors contributed equally as second authors

## 1. Introduction

Renal cysts are fluid-filled sacs or spaces that can pose a medical challenge, since they represent a heterogeneous group of malignant and benign lesions with overlapping clinical and radiologic features, and usually warrant further investigation[1]. They are usually detected using medical imaging, with an incidental detection rate of about 40% for patients who undergo abdominal computed tomography (CT) [2]. CT plays a major role in the clinical workflow for these patients, both in detection and in the management and follow-up of patients. In 1986, the Bosniak classification system was introduced to assess renal cysts based on CT scans to estimate malignancy risk[3]. While there have been incremental improvements over the decades, with e.g. the revision of the Bosniak classification system in 1997[4], 2005[5], and 2019[6], and the addition of other imaging modalities such as ultrasound and magnetic resonance imaging (MRI), the overtreatment of potentially benign cysts remains an issue, as aggressive surgical removal can cause harm to patients without clear benefit[7]. A method to improve overall patient outcomes is the development of artificial intelligence-based approaches to analyze and annotate clinical images. Such artificial intelligence (AI) models have the potential to increase detection and improve classification, as well as automatically annotate large datasets that can be used to train junior radiologists[8] and develop further image analysis systems. Cyst segmentation on medical images is not part of the clinical workflow as it takes a considerable amount of time and is usually subject to inter-observer variability[9–11]. Accurate segmentation of the lesions of interest could play an important role in improving the assessment of the malignancy risk and progression of the cysts, as well as improving the performance of downstream AI algorithms for classification and monitoring, where the lack of annotations is considered a bottleneck for quantitative studies[12].

Deep neural networks are becoming popular tools to aid in clinical detection, classification, and regression tasks, due to their rapid advances in fields outside of medicine, partially driven by increasingly available data and computational resources. A particular class of deep neural networks that can take 2D and 3D images as input are convolutional neural networks (CNNs), and they have demonstrated state-of-the-art performance in various medical image segmentation challenges [13] [14]. Despite their proven efficacy, CNNs operate as 'black boxes' due to their complex multi-layered architecture in which numerous filters interact non-linearly [15]. Ensuring the explainability of segmentation models is crucial not only for building trust and guaranteeing reliability but also for preventing model failures and fulfilling ethical and legal obligations by providing insights into the model's decision-making process [16]. Heatmaps, or attribution maps, are among the most common post-hoc explainability methods, as they highlight the regions of the input image that contribute most significantly to the model's decision [17]. The value of these maps for interpreting segmentation models is limited because they highlight a general area of relevance but cannot offer insights into the model's decisions on a pixel level [18]. Alternatively, counterfactual explanations enhance model understanding by simulating 'what-if' scenarios, where altering specific regions of an image can change the outcome[19]. This method aids in identifying the features that influence the segmentation model's decisions, achieved by modifying parts of the input image and observing the resultant changes in segmentation. This approach is interesting for medical image segmentation, as differences in segmented areas can affect decisions related to diagnosis and treatment, e.g. when monitoring changes in follow-up. Counterfactual explanations may potentially boost trust and reliability in the model's predictions by enabling users to understand causal relationships within the data[20]. Furthermore, quality control in medical image segmentation also improves trust, as data heterogeneity can lead to unnoticed segmentation errors that undermine the reliability of the results[21]. Uncertainty estimation can increase the reliability of segmentation methods by identifying failed segmentations for better quality control and indicating when human intervention is necessary[22]. Uncertainty estimation can also aid in interpreting counterfactual explanations by identifying unreliable parts of the generated images, thereby pinpointing features about which the model is ambiguous.

Another method to quantify regions of interest (ROIs) in medical images is the extraction of pre-defined quantitative features, including aspects such as texture, shape, and intensity, commonly known as handcrafted radiomics[23–25]. When a segmentation mask is available, radiomics features can be used to assess changes within the region of interest, making them particularly well-suited for interpreting segmentation networks[18]. Radiomics features can be useful in interpreting the changes within the region of interest that result in modifications to segmentation masks during the generation of counterfactuals.

In this study, we introduce an automated and explainable deep learning framework for the detection and 3D volumetric segmentation of renal cysts in contrast-enhanced abdominal CT scans, validated across a multi-centric dataset. We incorporate a novel counterfactual explanation method to provide insights into the framework's decision-making process by generating synthetic images. Additionally, we incorporate segmentation quality control by utilizing uncertainty estimation to identify potential inaccuracies. For a comprehensive analysis, we employ radiomics features for the quantitative interpretation of counterfactual explanations and apply uncertainty estimation for a thorough qualitative understanding of these explanations.

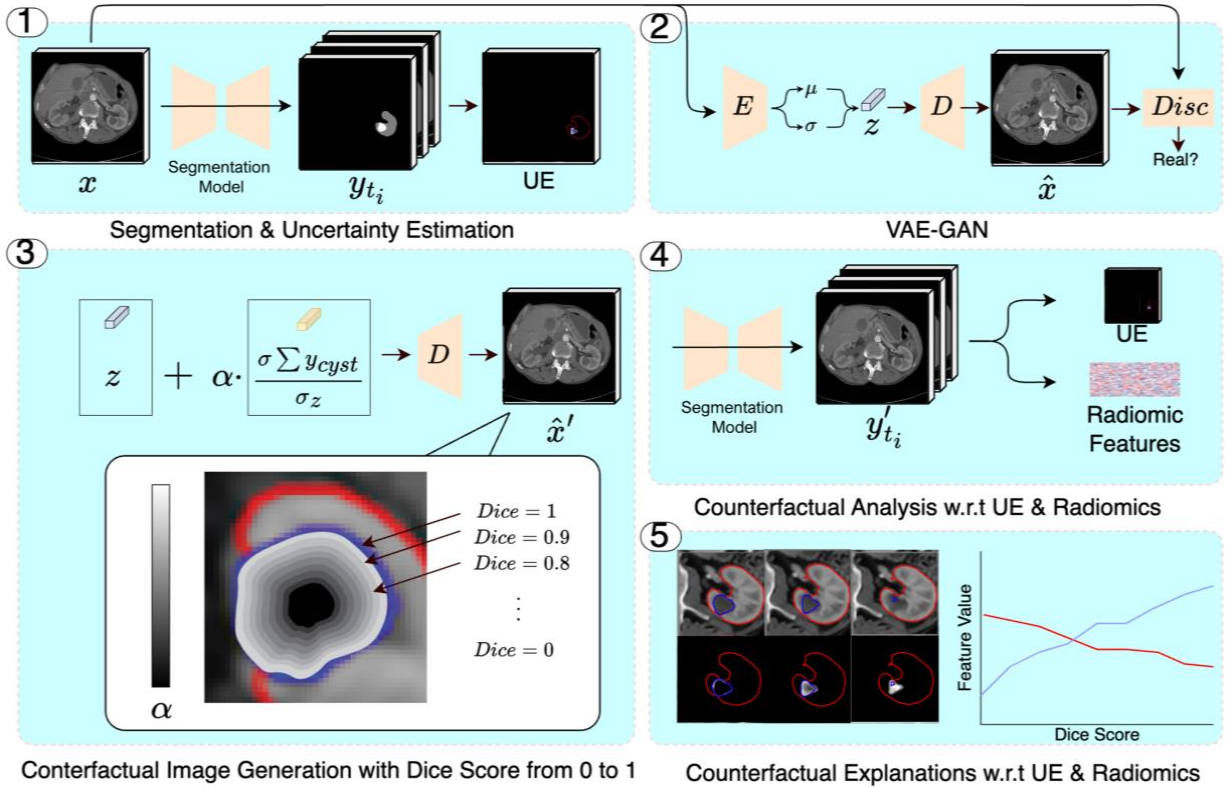

*Figure 1:* Proposed workflow for the segmentation of renal Cysts incorporating qualitative and quantitative evaluation of counterfactual explanations and uncertainty estimation. (1) The segmentation model generates a segmentation mask along with uncertainty estimates for the mask. (2) The original image is input into the VAE-GAN framework to produce a reconstructed image. (3) This reconstructed image is then fed into both the segmentation model and the VAE-GAN to generate segmentation masks and latent space representations, respectively. Latent space perturbations are created, leading to the production of new reconstructed images corresponding to segmentation masks with Dice similarity coefficient (DSC) ranging from 0 to 1. (4) These reconstructed images are re-input into the segmentation model to generate new segmentation masks, which serve as counterfactual explanations. At this point, uncertainty estimates and radiomics features are also extracted. (5) Counterfactuals and uncertainty estimation plots are employed to identify discrepancies in the model. Furthermore, the relationship between DSC and radiomics features is analyzed.

## 2. Methods
### 2.1 Description of the data

Contrast-enhanced CT scans (CE-CT scans) of 568 patients were retrospectively collected and anonymized by each center, with approval from the respective institutional review boards. Maastricht University Medical Center+ specifically waived the requirement for informed consent due to the retrospective, anonymized nature of the data collection, which involved no participant interventions or compensations. The segmentation of the regions of interest (ROIs) in the datasets from the University of California - San Francisco (UCSF) Medical Center, Maastricht University Medical Center (MUMC+), and Amsterdam University Medical Center (AMC) was conducted using MIM software (MIM Software Inc., Cleveland, OH) by a medical doctor with 4 years of experience in image segmentation and reviewed by a radiologist with 15 years of experience in medical radiology. These expert segmentations were established as the ground truth for both training and subsequent evaluation. To compile the images, radiology databases at UCSF, MUMC+, and AMC medical centers were searched using the keyword 'Bosniak'. Patients with radiologic reports indicating the presence of renal cyst(s) were identified. The inclusion criteria for this study were: (i) confirmation of a renal cyst and (ii) availability of an arterial phase contrast-enhanced CT scan. Additionally, an open-source dataset from the KITS 23 challenge was utilized, selecting only cases that contained renal cysts. The specifics of these datasets are detailed in Table 1.

### 2.2 Segmentation

In the initial phase, we utilized a low-resolution nnUNet for kidney and cyst segmentation to define the region of interest (ROI), ensuring focus on the kidney region during the final prediction [26]. CT images were resampled to the median resolution of $0.775 \times 0.775 \times 2.5 \ mm^3$ , and the intensity values were clipped to the 0.5th and 99.5th percentiles based on the training set, corresponding to values of -71 HU and 344 HU, respectively. Subsequently, z-score normalization was applied to the intensities, using a mean of 117.1 and a standard deviation of 89.9. A bounding box was created based on predictions from low-resolution nnUNet to isolate the ROI. We employed five-fold cross-validation to predict bounding boxes for cases within the training set. To address failure cases, the minimum size of the bounding boxes was set to match the median size of all predicted bounding boxes in the training set. This isolation of the ROI enabled the subsequent stages of cyst segmentation and counterfactual explainability to concentrate solely on the kidney region, thereby reducing the complexity of the models.

The segmentation network consisted of an adapted version of the Dynamic 3D UNet (DynUnet) implementation available in MONAI library[27]. For training, we ensured that 50% of the dataset includes some area beyond the bounding box to increase robustness. However, inference was performed only within the area of the bounding box predicted by the previous stage. The patch size was set to $128 \times 128 \times 48 \ pixels^3$. The augmentations applied included random zoom, random Gaussian noise, random Gaussian smoothing, random intensity scaling, and random flipping along three axes. All augmentations, except for random flipping, were applied with a probability of 0.15, while random flipping was applied with a probability of 0.5. The details about the augmentations are presented in Table A1. The loss function used for training the network combined multi-class dice loss[28] and binary cross-entropy loss with equal weight. We employed deep supervision in our training strategy and set the number of deep supervision layers to four [29,30]. In deep supervision, we aggregated the losses from the different layers of the U-Net and assigned decreasing weights to these layers, starting from the last layer which had the highest resolution. The weights decreased progressively, with 1.0 for the last layer, followed by 0.5, 0.25, 0.125, and finally 0.0625 for each preceding layer.

### 2.3 Uncertainty Estimation

Segmentation quality control can be enhanced by integrating an uncertainty estimation that gauges the model's confidence in the masks to mitigate silent failures. The estimation of the predictive uncertainty is facilitated by the posterior sampling of the weight space, achieved through inference from various models saved during stochastic gradient descent (SGD) training[31,32]. We employed a cyclic learning rate with a total training budget of $T_{max} = 1200$ epochs that are divided into 3 cycles. Each cycle consisted of $T_{cycle} = 400$ epochs, with the learning rate decreasing gradually until $80\%$ of $T_{cycle}$ had elapsed, after which it remained constant. The learning rate $lr$ at any point in $T_{cycle}$ is given by:

$$lr(t_{epoch}) = 0.01 \cdot \left[1 - \frac{min(t_{mod}, 320)}{400}\right]^{0.9}, t_{mod} > 0$$

Where $t_{mod} = t_{epoch} \% T_{cycle}$ represents the number of epochs elapsed within $T_{cycle}$. At the beginning of each cycle, the learning rate was increased to 0.1 for one epoch to escape the local minimum. After the training reaches a local minimum following 320 epochs within $T_{cycle}$, n=10 models are saved randomly from each cycle. The standard deviation of the predictions from these n=30 models across the 3 cycles offered a measure of the uncertainty associated with the predicted segmentation masks.

For the quality control, we considered segmentations with a Dice similarity coefficient (DSC) below 0.75 to be poor. To evaluate the effectiveness of using uncertainty in segmentation quality control, we plotted the proportion of images classified as having poor segmentation against the proportion of images marked for manual correction across various uncertainty thresholds [22,33]. Uncertainty estimation can be correlated with counterfactual synthetic images to identify characteristic features within the cyst and kidney that increase the model's uncertainty.

### 2.4 Counterfactual Explanation

Counterfactual explanations are produced by applying minimal alterations to the input image to change the model's prediction[15,34]. For segmentation networks, we create counterfactual images to change the predicted segmentation masks, providing insights into the image features that the model considers crucial for segmenting the region of interest. These explanations are generated using the variational autoencoder-generative adversarial network (VAE-GAN) framework, which establishes a latent space for image synthesis.

VAE-GAN merges the capabilities of a variational autoencoder (VAE) and a generative adversarial network (GAN), harnessing the GAN discriminator's learned feature representations to improve VAE's reconstruction and generative capabilities[35]. VAE consists of an encoder, $E$, that takes an input image $x$ to generate a latent representation $z$. The decoder, $D$, then decodes z to generate the reconstructed image, $\hat{x}$. VAE employs variational inference to learn a continuous latent representation with mean $\mu$ and standard deviation $\sigma$, assuming that z is from a Gaussian distribution. A GAN comprises two neural networks: the generator and the discriminator denoted as $Disc$. Within this framework, the decoder of the VAE functions as the generator, creating synthetic images. The discriminator is tasked with discerning between the training samples, $x$, and these generated images, $\hat{x}$, by assigning a probability ranging from 0 to 1. This process is conducted while simultaneously encouraging the VAE to generate data that closely aligns with the true data distribution. The training of the VAE-GAN utilizes a loss function defined as follows:

$$L_{total} = L_{recon} + L_{recon\_kid} + \lambda_{KL} * L_{KL} + \lambda_{perc} * L_{perc} + \lambda_{adv} * L_{adv}$$

Where $L_{recon} = \frac{1}{N}\sum_{i=1}^{N}|x - \hat{x}|$ is the reconstruction loss which measures the difference between the input image x and the reconstructed image $\hat{x}$, $L_{recon\_kid}$ calculates the reconstruction loss exclusively within the combined kidney and cyst region to ensure that the VAE-GAN framework learns the representative features within the ROI, $L_{KL} = D_{KL}(q_\varphi(z|x) \parallel p(z))$ is the Kullback-Leibler divergence loss which measures deviations of the learned latent variable distribution $q(z|x)$ from the prior distribution $p(z)$[36], $L_{perc} = \sum_{l=1}^{L} \frac{1}{M_l}|\phi_l(x) - \phi_l(\hat{x})|^2$ is the perceptual loss which measures the high-level perceptual and semantic differences between input $x$ and the reconstruction $\hat{x}$ using features extracted from a feature map $\phi_l(\cdot)$ obtained from layer $l$ and $M_l$ is the total number of elements in the feature map of layer $l$[37], and $L_{adv} = log(Disc(x)) + log(1 - Disc(\hat{x}))$ is the adversarial loss which measures the discriminator's ability to correctly classify real and reconstructed images. $\lambda_{KL} = 1 \times 10^{-6}$, $\lambda_{perc} = 0.001$, $\lambda_{adv} = 0.01$ are the hyperparameters that weigh the contribution of each corresponding loss component to the total loss. $L_{adv}$ is incorporated in the total loss function following 10 initial warm-up epochs. We used MONAI for implementing the VAE-GAN framework[27].

Counterfactuals are generated by perturbing the latent space of the VAE with respect to the gradients of the predicted segmentation mask. The process starts with reconstructing the input image using the VAE. This reconstructed image, which closely resembles the original, is then used to generate the counterfactuals. The region of interest corresponding to the ground truth cyst area is then selected to examine the impact of change of features within the ROI on the predicted segmentation masks. The predicted probabilities for cyst, kidney, and background classes are aggregated within this ROI. The latent space representation of the reconstructed image computed using the VAE is denoted by $z_{orig}$. The gradients of the aggregated probabilities of cyst, kidney, and background classes were computed separately with respect to $z_{orig}$, denoted by $\frac{\delta(\sum pred_{cyst})}{\delta z_{orig}}, \frac{\delta(\sum pred_{kid})}{\delta z_{orig}}$ and $\frac{\delta(\sum pred_{bg})}{\delta z_{orig}}$ respectively. A counterfactual image, $\widehat{x_{ci_\alpha}}$, was generated by the decoder as follows:

$$\widehat{x_{ci_\alpha}} = D\left(z_{orig} + \alpha(\frac{\delta(\sum pred_{cyst})}{\delta z_{orig}} - \frac{\delta\delta(\sum pred_{kid})}{\delta z_{orig}} - \frac{\delta(\sum pred_{bg})}{\delta z_{orig}})\right)$$

Where $\alpha$ determines the magnitude of increase or decrease of the cyst probabilities within the ROI, which in turn impacts the predicted cyst segmentation masks. The value $\alpha$ is iteratively increased and decreased to identify the upper and lower bounds until a point is reached where the consecutive monotonic increases or decreases in the DSC within the ROI no longer persist. The counterfactual images are generated with predicted DSC ranging from 0 to 1, in increments of 0.1.

Handcrafted radiomics features, including first-order statistics and texture features (Table A2), are extracted from counterfactual scans with varying Dice scores, using the ground truth ROI as the mask. These features are extracted using Pyradiomics (V3.1.0)[38] and the bin width is set at 25. These features provide insights into the internal density, heterogeneity, texture uniformity, and intensity distribution of renal cysts. They can be useful for investigating how changes in quantitative features within the ROI affect the predicted segmentation mask for the renal cyst. These handcrafted radiomics features are extracted from synthetic images that represent counterfactual explanations for the cysts in the test set, covering a range of Dice Similarity Coefficients (DSC) from 0 to 1. For each specific cyst, counterfactuals are generated for 10 distinct DSC values, each separated by an interval of approximately 0.1. For each case, a line is plotted with the DSC on the x-axis and the corresponding feature value on the y-axis. A median line is calculated from these individual lines, and the slope of this median line is also determined. The top three features associated with the largest positive slopes, and the top three associated with the largest negative slopes, are identified.

## 2.5 Evaluation Metrics

The performance of the kidney and cyst segmentation was evaluated using the Dice similarity coefficient $DSC$, which measures the overlap between the predicted segmentation mask and the ground truth mask. A **$DSC$** of greater than 0.10 was considered a detection, consistent with prior research that assessed detection capabilities for 3D lesions[39–41]. A candidate cyst is classified as a true positive $TP$ when its $DSC$ with the ground truth cyst exceeds 0.1. It is deemed a false positive $FP$ if this criterion is not met. A ground truth cyst lacking a corresponding candidate cyst prediction with a $DSC$ greater than 0.1 is labeled as a false negative $FN$. The detection performance is assessed using sensitivity and the number of false positives per image.

The counterfactual generation capabilities are directly correlated with the reconstruction ability of the VAE-GAN. We propose a counterfactual reliability index (CRI) that compares the **$DSC$** of the original images ($DSC_{orig}$) with $DSC$ of the reconstructed images ($DSC_{recon}$), as well as the sensitivity of the original images ($Sensitivity_{orig}$) with the sensitivity of the reconstructed images ($Sensitivity_{recon}$). This metric quantifies the reliability level of the counterfactuals by measuring how closely the performance of the reconstructed images correlates with that of the original images. To achieve a high CRI close to 1, the reconstructed images must exhibit segmentation accuracy and detection capability similar to those of the original images. The CRI is defined as follows:

$$CRI = \left(1 - \frac{\left|DSC_{orig} - DSC_{recon}\right|}{DSC_{orig}}\right) \times \left(1 - \frac{\left|Sensitivity_{orig} - Sensitivity_{recon}\right|}{Sensitivity_{orig}}\right)$$

*Table 1:* Description of the datasets used in this study.

| No | Name of the dataset | Use | Medical Center | No. of CT Scans | CT Phase | No. of Cysts | No. of Malignant Cysts | Mean Cyst Volume (ml) | Median Resolution ($mm^3$) |
|---|---|---|---|---|---|---|---|---|---|
| 1. | UCSF | Training/ Testing | University of California - San Francisco | 217 | Late Arterial Phase | 515 | 17 | 44.8 | $0.76 \times 0.76 \times 2.4$ |
| 2. | KITS2023 | Training/ Testing | Open Source | 248 | Late Arterial Phase and Nephrogenic Contrast Phase | 748 | NA | 47.2 | $0.81 \times 0.81 \times 3.38$ |
| 3. | MUMC | External Validation | Maastricht University Medical Center | 60 | Late Arterial Phase | 115 | 15 | 59.5 | $0.74 \times 0.74 \times 2.32$ |
| 4. | AMC | External Validation | Amsterdam University Medical Center | 43 | Late Arterial Phase | 108 | 9 | 46.4 | $0.71 \times 0.71 \times 2.26$ |

*Table 2:* Overview of the quantitative model performance.

| Dataset | No. of CT Scans | No. of Cysts | DSC | DSC Small Cysts | DSC medium Cysts | DSC large | Sensitivity | Sensitivity small cysts | Sensitivity medium cysts | Sensitivi ty large cysts | FPPI |
|---|---|---|---|---|---|---|---|---|---|---|---|
| **Cross-Validation Results** | | | | | | | | | | | |
| UCSF | 217 | 515 | 0.83 | 0.63 | 0.75 | 0.90 | 0.96 | 0.89 | 0.97 | 1.0 | 0.47 |
| KITS | 248 | 748 | 0.83 | 0.63 | 0.75 | 0.90 | 0.94 | 0.87 | 0.94 | 1.0 | 0.25 |
| UCSF Reconstructed | 217 | 515 | 0.81 | 0.57 | 0.79 | 0.91 | 0.93 | 0.83 | 0.97 | 1.0 | 0.40 |
| KITS Reconstructed | 248 | 748 | 0.82 | 0.58 | 0.73 | 0.88 | 0.92 | 0.83 | 0.94 | 1.0 | 0.32 |
| **External Test Set Results** | | | | | | | | | | | |
| AMC | 43 | 115 | 0.83 | 0.65 | 0.82 | 0.90 | 0.95 | 0.88 | 1.0 | 1.0 | 0.16 |
| MUMC | 60 | 108 | 0.84 | 0.63 | 0.82 | 0.89 | 0.97 | 0.92 | 1.0 | 0.97 | 0.25 |
| AMC Reconstructed | 43 | 115 | 0.82 | 0.58 | 0.82 | 0.90 | 0.94 | 0.84 | 1.0 | 1.0 | 0.14 |
| MUMC Reconstructed | 60 | 108 | 0.82 | 0.59 | 0.81 | 0.87 | 0.94 | 0.83 | 1.0 | 0.97 | 0.21 |

## 3. Results

A total of 568 contrast-enhanced CT scans containing 1486 renal cysts were employed for training, testing, and external validation of the detection and segmentation algorithm, as well as for validating counterfactual generation and uncertainty-based quality control measures. Five-fold cross-validation was conducted on the KITS23 and UCSF datasets. Datasets from two centers, AMC and MUMC, served as external test datasets. The specifics of these datasets are detailed in Table 1.

### 3.1 Segmentation

We classified renal cysts into three categories based on their percentile distribution within the datasets. Cysts with a volume of less than *1.8 ml* were considered small. Those with volumes ranging from *1.8 ml* to *11.5 ml* were categorized as medium. Cysts with a volume greater than *11.5 ml* were classified as large. We evaluated the segmentation model's performance in terms of the DSC and sensitivity across various cyst sizes, as well as the false positives per image (FPPI) for both original and reconstructed scans. Table 2 presents the model's quantitative performance on original and reconstructed images for both the five-fold cross-validation and the test set.

During the five-fold cross-validation of the UCSF and KITS datasets, a mean DSC of 0.83 was achieved for both datasets. For the reconstructed images from the VAE for generating counterfactual explanations, the mean DSC was 0.81 for the UCSF dataset and 0.82 for the KITS dataset, respectively. Fig. 2(a) illustrates the distribution of the DSC during five-fold cross-validation for original and reconstructed images, as well as for cysts of various sizes. There was no significant difference in the overall DSC between the original and reconstructed images. The detection sensitivity for the UCSF and KITS datasets was 0.96 and 0.94, respectively. For the reconstructed dataset, the sensitivity for the UCSF and KITS datasets was 0.93 and 0.92, respectively. During the five-fold cross-validation, all malignant cysts were detected in both the original and reconstructed images, with their Dice distribution shown in Fig. 2(c). There was no significant difference in the DSC for medium and large cysts between the original and reconstructed images. However, there was a significant difference in the DSC for small cysts, with scores of 0.63 for both the UCSF and KITS datasets and scores of 0.57 and 0.58 for the reconstructed images of the UCSF and KITS datasets, respectively. The counterfactual reliability index for the UCSF and KITS datasets was 0.95 and 0.97, respectively.

There was no significant difference in the test set in terms of the DSC between the original and reconstructed images for the overall DSC and the DSC of small, medium, and large cysts. Fig. 2(b) illustrates the distribution of DSC on the test set for overall performance as well as for cysts of various sizes. On the AMC and MUMC datasets, a DSC of 0.83 and 0.84 were observed for the original images, respectively, and a DSC of 0.82 was recorded for both datasets for the reconstructed images. The detection rate for the AMC and MUMC datasets showed an overall sensitivity rate of 0.95 and 0.97 for the test images, respectively, and a detection rate of 0.94 for both datasets for the reconstructed images. For medium and large cysts, sensitivities of 1 and 0.97 were observed, respectively, for both the original and reconstructed images. For small cysts, the sensitivity for the AMC and MUMC datasets was 0.88 and 0.92 for the original images, respectively, and a comparatively lower sensitivity of 0.84 and 0.83 for the reconstructed images. A sensitivity of 0.96 was observed for malignant cysts for both original and reconstructed images, with their DSC shown in Fig. 2(d). Additionally, the original images exhibited a higher false positive rate per image of 0.16 and 0.25 for the AMC and MUMC datasets, respectively, compared to 0.14 and 0.24 for the reconstructed images. The counterfactual reliability index for the AMC and MUMC datasets was 0.97 and 0.95, respectively. This demonstrates that the reconstructed datasets produce segmentation and detection results very similar to those of the original datasets. Consequently, these reconstructed images can be used to generate counterfactuals that accurately reflect the model's true behavior in comparison to the original images.

## 3.2 Counterfactual Explanation

In our study, we used gradient manipulations of the predicted segmentation mask and VAE-GAN to generate counterfactual images with the intent of altering the model's predicted cysts segmentation masks. The counterfactual explanations present insights into the interpretability and robustness of the segmentation model to intensity variations within the cyst region. Fig. 3 provides a detailed visualization of the impact of image feature manipulations on cyst segmentation performance using counterfactual generation on the test set cases. These counterfactual variations help identify the features that the segmentation model deems significant, such as textures, and internal cyst densities, by observing how manipulating these features influences the DSC. The orange box highlights the baseline original images which serve as a reference for subsequent counterfactual perturbations. In Fig. 3 (a), a gradient of texture modifications is observed, demonstrating that as the texture of the cysts becomes more similar to that of the kidney, the segmentation model increasingly classifies these areas as kidney tissue, which is reflected in the decreasing DSCs. Conversely, enhancing the texture contrast leads to an improvement in the DSCs, suggesting that the model relies heavily on texture differentiation between cysts and kidney tissue for accurate segmentation. Fig. 3 (b) shows a scenario where the model's performance is highly sensitive to even minimal perturbations in the image features. Despite the lack of significant changes in the cysts' intensity, there is a pronounced effect on the segmentation output, indicating that the model's accuracy is finely tuned to specific image characteristics. Fig. 3 (c) shows the impact of perturbations for counterfactual generation for a small cyst. The accompanying DSCs serve as a quantitative measure of segmentation accuracy that correlates directly to the visual changes within the cyst.

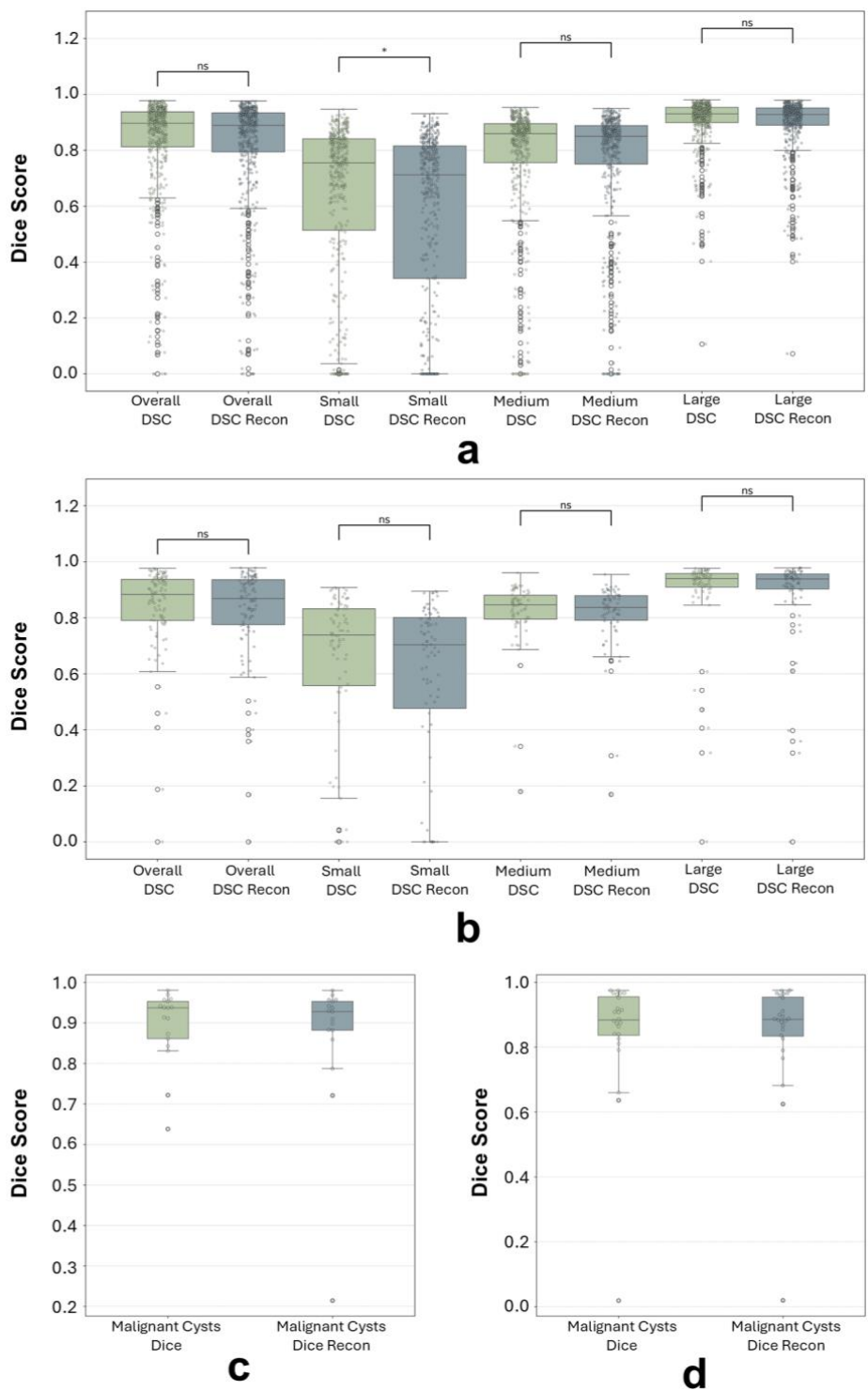

*Figure 2:* Quantitative evaluation of segmentation efficacy by dice coefficient, stratified by cyst volume and differentiated by standard and reconstructed images from VAE-GAN for counterfactual analysis. Statistical significance was determined using the two-sided Mann–Whitney–Wilcoxon test, where "ns" indicates $p > 0.05$, and "*" denotes $p < 0.05$. (a) displays five-fold cross-validation results, detailing the Dice Coefficient for overall patients, as well as for small ($< 1.8\,ml$), medium ($1.8 - 11.5\,ml$), and large cysts ($> 11.5ml$), with respective p-values for comparisons between standard and reconstructed methods being 0.17, <0.001, 0.16, and 0.17. (b) presents the test set Dice coefficient results for both standard and reconstructed methods, yielding p-values of 0.60, 0.12, 0.73, and 0.68, respectively. (c) compares the performance of the standard and reconstructed methods based on the Dice coefficient for malignant cysts in five-fold cross-validation, while (d) shows their performance on the test set for malignant cysts. (For high resolution: Link)

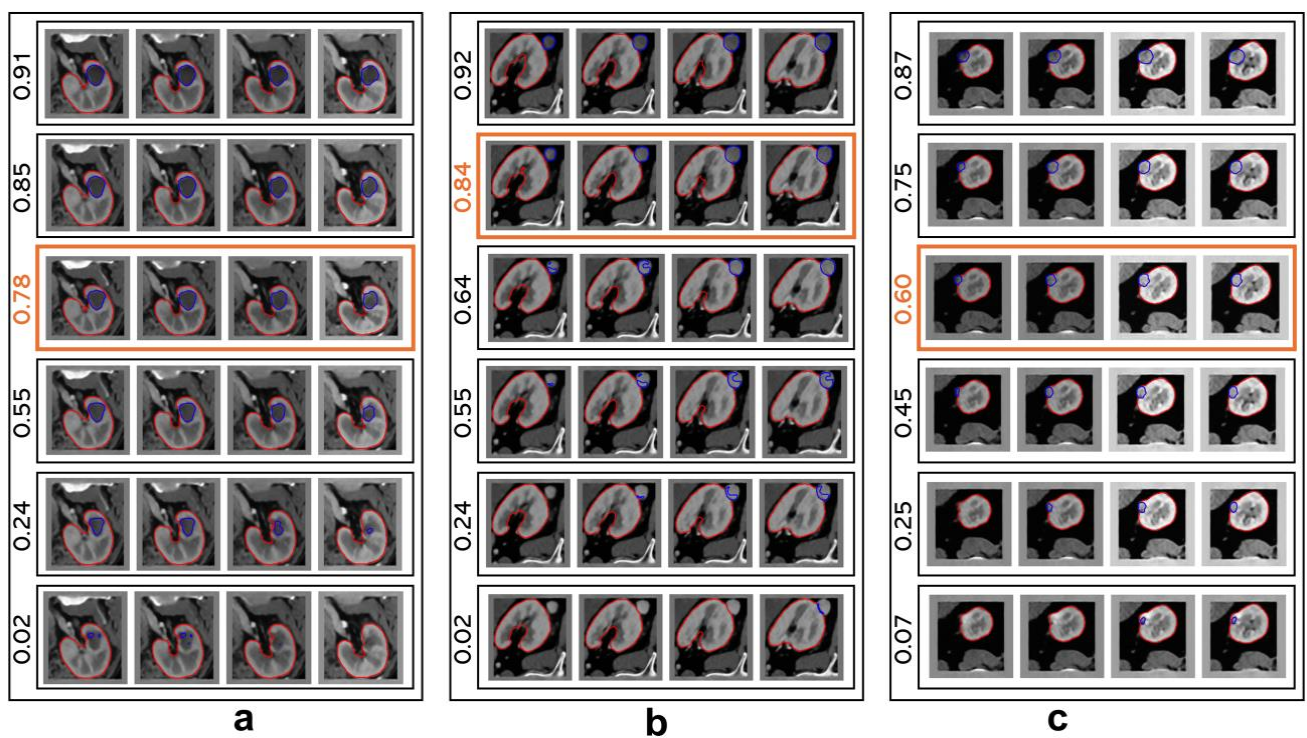

*Figure 3:* Counterfactual explanations for cyst segmentation accuracy in a segmentation network. (a), (b), and (c) illustrate the segmentation network's performance, with the Dice similarity coefficient (DSC) for the cyst displayed next to each image, demonstrating the accuracy of cyst segmentation. During counterfactual image generation, only the attributes of the cyst being investigated are altered, directly influencing the displayed DSC. The network's segmentations are encapsulated within orange boxes, where the red boundary delineates the kidney and the blue boundary defines the cyst. These images display a sequence of counterfactual modifications that result in both increased and decreased DSC, highlighting how specific changes to the cyst's representation affect segmentation accuracy. (For High Resolution: Link)

An analysis was conducted to explore if there is a meaningful correlation between various radiomics features and the DSCs of generated counterfactuals. Fig. 4 shows the relationship between the DSCs of the generated counterfactuals and the observed changes in different radiomics features within the ground truth region. A grey line in the background represents a plot between the DSC and the corresponding value of a radiomics feature for an image in the test set. The colored lines show the average of these lines along with the confidence interval. The top three features exhibiting the highest positive slopes of the average lines are indicated in the top row (a-c) of Fig. 4. These features are: Original Gray Level Dependence Matrix (GLDM) LowGrayLevelEmphasis, Original Gray Level Co-occurrence Matrix (GLCM) MaximumProbability, and Original GLCM JointEnergy. A higher value of Original GLDM LowGrayLevelEmphasis suggests a predominance of low-intensity pixels. Original GLCM MaximumProbability reflects the likelihood of pairs of pixels having the same value, with higher scores indicating more consistent texture patterns. Original GLCM JointEnergy represents the sum of squared elements in the GLCM matrix, with increasing values demonstrating more homogeneity. Conversely, the features showing the top three highest negative slopes are displayed in the middle row (d-f) of Fig. 4. These features are: Original Neighborhood Gray Tone Difference Matrix (NGTDM) Contrast, Original GLCM JointAverage, and Original GLCM SumAverage. Original NGTDM Contrast measures local variations in gray level intensity, with a higher value indicating greater heterogeneity within the cyst. Original GLCM JointAverage calculates the average intensity of pairs of adjacent pixels, providing an idea of the general level of gray intensity, and Original GLCM SumAverage sums the average intensities of pixel pairs, offering insight into the overall intensity distribution. Lastly, the bottom row (g-i) shows plots for Original Gray Level Size Zone Matrix (GLSZM)

LargeAreaHighGrayLevelEmphasis, Original GLSZM Zone Variance, and Original GLSZM LargeAreaEmphasis, which do not display a significant trend with variation in DSCs as indicated by a flat average line.

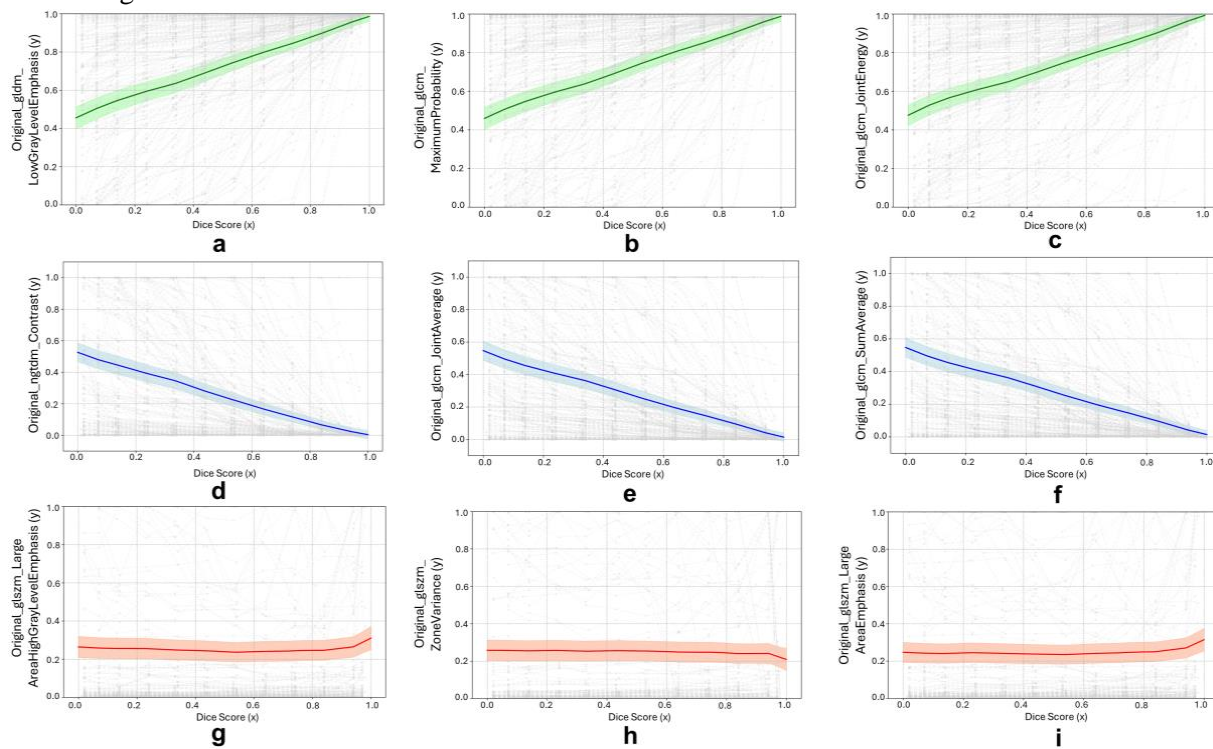

*Figure 4:* Correlation between Dice similarity coefficient (DSC) variations and radiomic feature alterations via counterfactuals on the test set. This figure illustrates the relationship between changes in DSCs and their impact on radiomics features extracted from the region of interest—the ground truth—across counterfactual scenarios. DSCs range from 0 to 1. The top row (a-c) displays features 'Original GLDM LowGrayLevelEmphasis,' 'Original GLCM MaximumProbability,' and 'Original GLCM JointEnergy', which exhibit a positive correlation with increasing DSCs. The middle row (d-f) depicts features 'Original NGTDM Contrast,' 'Original GLCM JointAverage,' and 'Original GLCM SumAverage,' which show a negative correlation with decreasing DSCs. The bottom row (g-i) presents features 'Original GLSZM LargeAreaHighGrayLevelEmphasis,' 'Original GLSZM Zone Variance,' and 'Original GLSZM LargeAreaEmphasis,' that demonstrate no significant relationship with the DSC, highlighting the heterogeneity of radiomics feature behaviors in response to segmentation accuracy variations. Each plot includes individual case trends as light grey lines in the background, against which the aggregated trend for the test set is overlaid, providing both an overview and a granular look at the data. (For High Resolution: Link)

### 3.3 Uncertainty Estimation

Model confidence in predicted masks is assessed using posterior weight sampling to avoid silent failures by flagging uncertain cases for manual review. Segmentations with a DSC below 0.75 are deemed poor quality. Fig. 5 illustrates the application of uncertainty estimation for segmentation quality control in training and test sets. It plots the fraction of segmentations flagged for manual correction against the remaining poor segmentations fraction for various uncertainty thresholds. The dashed black line shows the trend for random flagging. The ideal trend is marked by the gray line and shaded area, indicating a proportional reduction of poor segmentations with increased manual correction. Fig. 5 (a) displays the segmentation quality control performance for the training set during five-fold cross-validation with 95% corresponding confidence intervals. Initially, 24% of segmentations are classified as poor (DSC < 0.75). Flagging 20% of the data for manual correction reduces the fraction of poor segmentations to 7%, indicating a 71% decrease. Fig. 5 (b) presents similar results for the test set. Initially, 17% of segmentations are poor, which falls to 5% upon flagging, a reduction of 69%. These findings confirm the effectiveness of uncertainty estimation as a means of enhancing the quality control process.

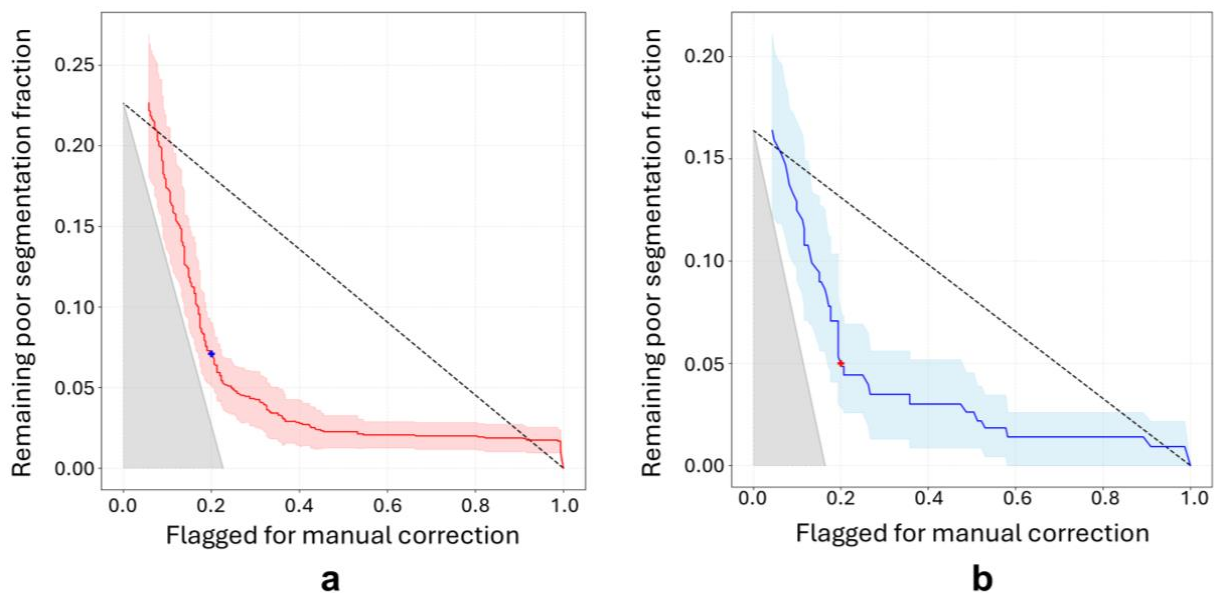

*Figure 5:* Segmentation quality control using uncertainty estimation for training and test sets. In both, the proportion of segmentations that remain after being flagged for manual correction is plotted against the threshold for flagging. The dashed black line represents the behavior for random flagging, where there is no information in the uncertainty estimates and samples are randomly chosen for manual correction, with no correlation between poor segmentation masks and flagging for manual correction. The ideal scenario is indicated by the gray-shaded area. Segmentations with a Dice Similarity Coefficient (DSC) below 0.75 are classified as poor quality. The solid colored lines track the actual data, with the training set performance in red and the test set in blue, facilitating direct comparison with the ideal and random flagging baselines. The red and blue markers denote the fraction of poor-quality segmentations left when only 20 percent of the dataset is retained for manual correction. (For High Resolution: Link)

We integrated counterfactual explanations with uncertainty estimation to enhance understanding of the segmentation model's reliability and to identify image features that contribute most to high segmentation uncertainty. Fig. 6 illustrates the use of counterfactual explanations and uncertainty estimation to analyze the model's behavior in test set cases. In Fig. 6(a), there is high initial uncertainty for the predicted cyst. When counterfactuals are generated with decreasing DSCs, uncertainty decreases until the cyst merges indistinguishably with the kidney at lower DSCs. Fig. 6(b) depicts a segmentation of a renal cyst with low initial uncertainty. However, as counterfactuals modify the image features, uncertainty increases. Even when the DSC of counterfactual drops to 0.07, the image retains some cyst-like characteristics but is segmented as kidney tissue, accompanied by increased uncertainty. Fig. 6(c) presents a predicted cyst experiencing minimal changes in image features during counterfactual generation, yet the DSC drops to zero and uncertainty increases. This indicates that even minor perturbations can lead to significant model uncertainty. The combined counterfactual explanations and uncertainty plots can be stitched together to form a Graphics Interchange Format (GIF), an example of which can be seen in Figure A1.

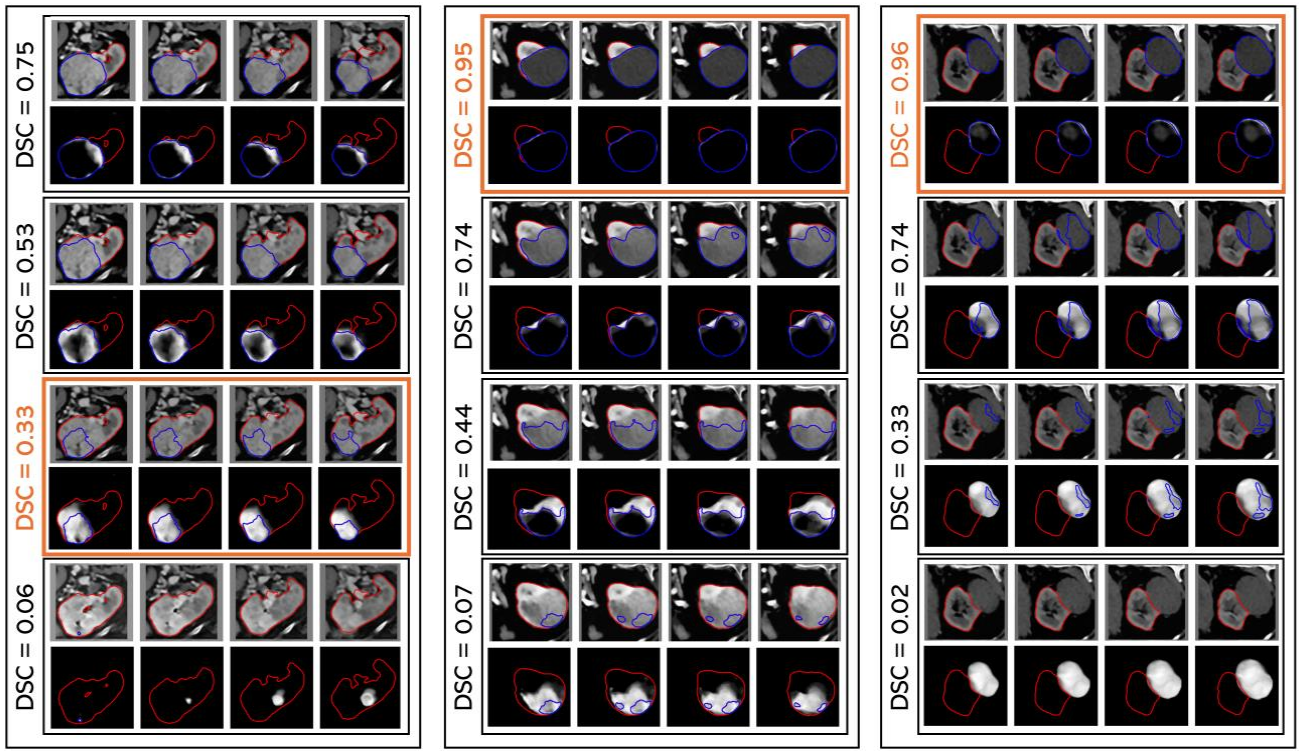

*Figure 6:* Integration of counterfactuals with uncertainty in segmentation evaluation. For each column labeled (a), (b), and (c), the orange box displays the original image, the model's segmentation output, and the corresponding uncertainty. The black boxes in each column correspond to counterfactual images that demonstrate a range of DSC values, showing how variations in image features can lead to different segmentation accuracies. The greater intensity of white in the uncertainty plot indicates higher uncertainty, denoting areas where the model's predictions are less confident. The DSC is noted on the left of each box, with the red and blue outlines depicting the kidney and cyst, respectively. These counterfactual images, paired with the uncertainty visualization, underscore the model's predictive variability and aid in the assessment of cyst characteristics that result in increased uncertainty. (For High Resolution: Link)

## 4. Discussion

Automated detection and segmentation of kidney cysts may significantly enhance the evaluation of malignancy risk and cyst progression, alongside boosting the efficacy of subsequent artificial intelligence (AI) algorithms used for classification and monitoring. The scarcity of annotations is identified as a critical limitation for quantitative studies, underscoring the importance of accurate lesion segmentation. Hence, we introduced a deep learning framework for 3D volumetric segmentation of renal cysts in CT scans, which incorporates uncertainty-based quality control and novel counterfactual explanations for segmentation. Despite the high performance attributed to convolutional neural networks (CNNs), they operate as black boxes, failing to explain how changes in features within the region of interest would impact the prediction. To gain insights into the model's decision-making process, we employed a VAE-GAN framework and manipulated the gradient of the segmentation to generate counterfactuals. We conducted both quantitative and qualitative evaluations of the proposed counterfactual explanations on a multi-centric dataset to validate the segmentation framework. This work is among the initial attempts to investigate how features within a specific region of interest in a 3D image, such as a cyst, impact the performance of the segmentation model while aiming to keep the remainder of the image constant.

The proposed segmentation model achieved DSCs of 0.83 and 0.82 for renal cysts on external test datasets, and cyst sensitivity of 0.95 and 0.97, respectively. Lin et. al[42] utilized a 3D-UNet to segment renal cysts in the corticomedullary phase of CT urography scans, obtaining a DSC of 0.54 on their test set. Blau et. al[43] used a 2D-UNet for the segmentation of renal cysts in abdominal CT and achieved a sensitivity of 0.85 on their test set. No significant difference in segmentation performance, as measured by the DSC, was observed between the original images and those reconstructed from the VAE for generating counterfactual explanations in the external test sets. The reconstructed images resulted in fewer false positives per image for both test datasets, which may be attributed to the noise-reduction capabilities of the VAE[44]. A sensitivity of 0.96 was achieved for malignant cysts in the test dataset, indicating that the segmentation masks predicted by our model could be utilized in future cyst characterization algorithms. The counterfactual reliability index on the test datasets was 0.97 and 0.94, indicating that the reconstructed images produced segmentations very similar to those of the original images.

The qualitative evaluation of the segmentation counterfactuals reveals that gradient-based manipulation of the disentangled latent space of the Variational Autoencoder (VAE) produces realistic counterfactuals. These may blend with the kidney or the background as the DSC approaches zero. However, even minimal perturbations within the cyst that do not result in significant visual changes can markedly affect the DSC. While segmentation counterfactuals can help visualize changes in image features, previous work has been mostly limited to attribution maps, which provide a limited intuitive understanding of the segmentation model[16]. Since the radiomic features were only extracted from the ground truth masks on the original and reconstructed scans, we were able to assess the relationship between DSCs and radiomics features within the cysts. The analysis revealed that a predominance of low gray level values, consistent texture, and homogeneity has a positive correlation with segmentation consistency, i.e higher DSCs between the ground truth segmentation and the model output, while heterogeneity, contrast, and overall brightness have a negative correlation with DSCs.

Uncertainty estimation in the proposed segmentation model demonstrated the ability to circumvent model failures by flagging them for manual correction. The joint analysis of uncertainty estimation and counterfactual visualizations provided insights into how changes in the image features of renal cysts can cause variations in model uncertainty. Even minimal modifications in the image features of the cyst, which are not visually perceptible, can result in increased model uncertainty. Furthermore, it also helps us understand how the image features need to change for the model uncertainty to decrease. Therefore, the combined plots are useful for understanding the model's confidence, reliability, and decision-making process in terms of image features. Counterfactuals with image features that cause high uncertainty can be included during model training to enhance the robustness of the segmentation model.

We conducted a comprehensive qualitative and quantitative analysis of our explainable segmentation framework but it was not without limitations. While we successfully generated counterfactuals with varying DSCs for regions that were successfully detected and analyzed regions associated with false positives, creating counterfactuals for false negatives proved challenging occasionally due to the limited magnitude of gradients in those regions of interest. Future research could concentrate on the counterfactual analysis of false negatives, particularly on overcoming the challenge of limited gradients. Additionally, the potential benefits of this research could be further investigated through an *in silico* trial that could assess whether the provision of explanations boosts clinicians' confidence and their performance in the detection and management of cysts[45]. Future research could also focus on the use of counterfactuals with model discrepancies for augmentation during model training to enhance the performance and reliability of segmentation models. Furthermore, in the future, this segmentation framework will be integrated into an automated malignancy classification algorithm for renal cysts, aiming to reduce unnecessary surgeries.

## 5. Conclusion

In this multi-center study, we introduced an innovative deep learning framework for the automated detection and segmentation of renal cysts in computed tomography (CT) scans that incorporates counterfactual explanations and uncertainty-based quality control. Our approach leverages the strengths of convolutional neural networks for segmentation while addressing their inherent 'black box' nature. The use of counterfactual images and uncertainty estimation provided insights into the algorithm's decision-making process by allowing us to observe changes in image features that lead to increases or decreases in the Dice similarity coefficient. Our comprehensive analysis across multiple datasets, including malignant cases, validates the effectiveness of our methodology in producing segmentation masks for renal cysts. These masks can be used in clinical practice and by algorithms designed for the classification of renal cyst malignancy.

## 6. Acknowledgements

Authors acknowledge financial support from the European Union's Horizon research and innovation programme under grant agreement: ImmunoSABR n° 733008, CHAIMELEON n° 952172, EuCanImage n° 952103, IMI-OPTIMA n° 101034347, RADIOVAL (HORIZON-HLTH-2021-DISEASE-04-04) n°101057699, EUCAIM (DIGITAL-2022-CLOUD-AI-02) n°101100633, GLIOMATCH n° 101136670, AIDAVA (HORIZON-HLTH-2021-TOOL-06) n°101057062, REALM (HORIZON-HLTH-2022-TOOL-11) n° 101095435. This work was partially supported by the Dutch Cancer Society (KWF Kankerbestrijding), project number 14449.

## 7. Disclosures

PL,HW: No disclosures related to the current manuscript. Outside of the current manuscript: PL has received grants and sponsored research agreements from Radiomics SA, Convert Pharmaceuticals SA, and LivingMed Biotech srl. He received presenter fees (in cash or in kind) and/or reimbursement of travel costs or consultancy fees (in cash or in kind) from AstraZeneca, BHV srl, and Roche. PL holds or held minority shares in Radiomics SA, Convert Pharmaceuticals SA, Comunicare SA, LivingMed Biotech srl, and Bactam srl. PL is a co-inventor on two issued patents with royalties on radiomics (PCT/NL2014/050248 and PCT/NL2014/050728), licensed to Radiomics SA. One issued patent on mtDNA (PCT/EP2014/059089), licensed to ptTheragnostic/DNAmito. One granted patent on LSRT (PCT/P126537PC00, US Patent No. 12,102,842), licensed to Varian. One issued patent on the radiomic signature of hypoxia (U.S. Patent No. 11,972,867), licensed to a commercial entity. One issued patent on prodrugs (WO2019EP64112) without royalties. One non-issued, non-licensed patent on deep learning-radiomics (N2024889). Three non-patented inventions (software), licensed to ptTheragnostic/DNAmito, Radiomics SA, and Health Innovation Ventures. PL confirms that none of the above entities were involved in the preparation of this paper. HW has minority shares in the company Radiomics SA.

# Appendix

*Table A1*: Overview of image data augmentation applied during training.

| Transformation Name | Parameters and Settings | Probability |
|---|---|---|
| Random Zoom | Zoom range: 0.9 to 1.2 (multiplicative scale factor, no units) | 0.15 |
| Random Gaussian Noise Addition | Gaussian noise standard deviation: 0.01 (relative intensity units) | 0.15 |
| Random Gaussian Smoothing | Gaussian smoothing sigma range: 0.5 to 1.15 pixels for all axes | 0.15 |
| Random Intensity Scaling | Intensity scaling factor: 0.3 (multiplicative factor, no units) | 0.15 |
| Random Flip Along Axis 0 | Flip along spatial axis 0 | 0.5 |
| Random Flip Along Axis 1 | Flip along spatial axis 1 | 0.5 |
| Random Flip Along Axis 2 | Flip along spatial axis 2 | 0.5 |

*Table A2:* Details about the handcrafted radiomics features used to analyze counterfactual images, which correspond to varying Dice scores ranging from 0 to 1, with a fixed region of interest based on the ground truth cyst annotation.

| Handcrafted Radiomics Features | Number of Features |
|---|---|
| First Order Statistics | 18 |
| Gray Level Co-occurrence Matrix (GLCM) | 24 |
| Gray Level Dependence Matrix (GLDM) | 14 |
| Gray Level Size Zone Matrix (GLSZM) | 16 |
| Neighboring Gray Tone Difference Matrix (NGTDM) | 5 |

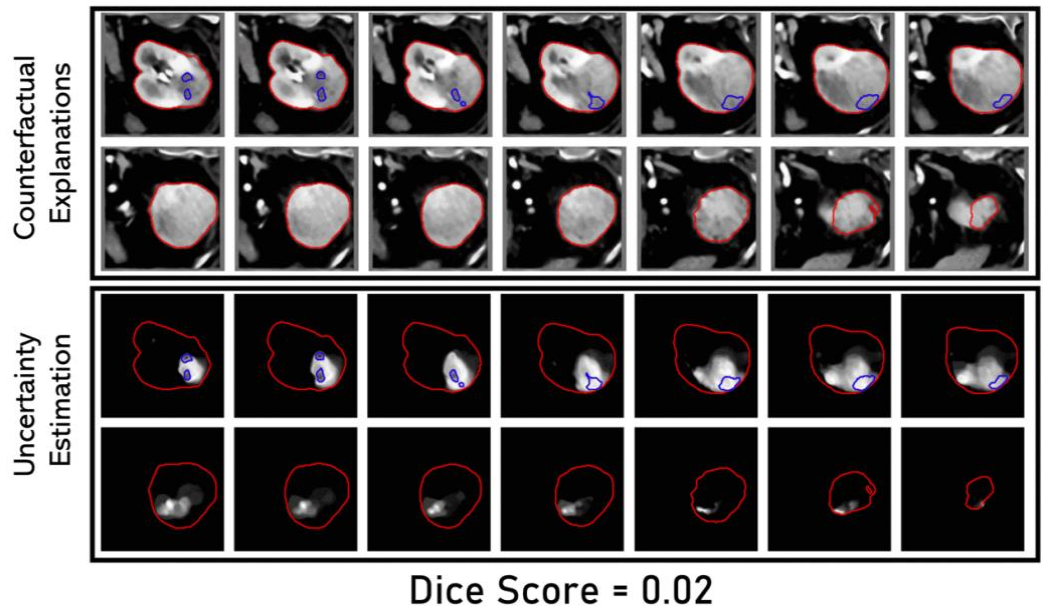

*Figure A1*: The counterfactual explanations, which are synthetic images produced by altering features in the cyst region of interest to generate predicted masks with varying Dice scores, are stitched together to form a GIF. The red contour corresponds to the kidney, and the blue contour corresponds to the predicted cyst. The uncertainty maps highlight the image features about which the model is uncertain. The GIF can be viewed at the following link: https://drive.google.com/file/d/1W_Q3YVn1y2U6sU2RiXNqqM7WtMk-QG3q/view?usp=sharing